\documentclass[12pt]{article} 
\usepackage{amssymb,latexsym,amsmath}
\usepackage{graphicx,times} 
\newtheorem{The}{Theorem}[section]
\newtheorem{Def}{Definition}[section]
\newtheorem{lem}{Lemma}[section]
\newtheorem{Ex}{Example}[section]
\newtheorem{Prop}{Proposition}[section]
\numberwithin{equation}{section}
\begin{document}
\begin{center}
{\LARGE {\bf Local Fractional Calculus: a Review}} 
\vskip 1cm
{\Large {\bf Kiran M. Kolwankar}}
\vskip 1cm
{\large {\bf Department of Physics, \\
Ramniranjan Jhunjhunwala College, \\
Ghtakopar(W), Mumbai 400086\, India}}\\
\medskip 
Kiran.Kolwankar@gmail.com\\
\end{center}

\begin{abstract}
The purpose of this article is to review the developments related to the notion of local fractional
derivative introduced in 1996. We consider its definition, properties, implications and possible
applications. This involves the local fractional Taylor expansion, Leibnitz rule, chain rule, etc.
Among applications we consider the local fractional diffusion equation for fractal time processes
and the relation between stress and strain for fractal media. Finally, we indicate a stochastic 
version of local fractional differential equation.
\end{abstract}
\vskip 1cm
\noindent

\section{Introduction}
In 1970's Mandelbrot~\cite{1Man} popularized fractals, irregular sets whose appropriately defined dimension is larger 
than their topological dimension, which were already known in the mathematical literature as pathological
examples. He advocated their use in modelling numerous irregular objects and processes found in nature.
This has completely changed the outlook and since then several studies have been carried out to determine
the fractal dimensions in diverse fields ranging from biology to astrophysics. A natural extension of these
developments leads us to related questions which involve fractals, for example, diffusion on fractals, wave
equation with fractal boundary condition etc. As a consequence, it becomes necessary to be able to incorporate
fractals in usual calculus or its appropriate generalization. However, fractals being generally non-differentiable
in some sense or the other, the ordinary calculus fails to apply.

A possible candidate to overcome this impasse is the fractional calculus~\cite{2OS,SAM,POD,Das,WBG,Hil}, a generalization of differentiation
and integration to arbitrary orders. The field which existed for quite long but being used in applications
only recently. The fractional derivative operators, as defined in the literature
using various approaches, turn out to be non-local operators and quite well incorporate and describe long
term memory effects and asymptotic scaling. They were also used to describe fractals because of the common
scaling property. It is known that fractals have a local scaling property and in general the scaling
exponent can be different at different places. Therefore, in author's thesis~\cite{KMK} and related publications~\cite{2KG1,2KG2,2KG3},
 it was thought to be prudent to suitably modify
the definition and make it local. This gave rise to the definition of the \emph{local fractional derivative}
(LFD). Several authors~\cite{LDE,KTAR,CC,CYZ,Wu,Yan,AC,BDG} have tried to take further and apply this definition. Some of these works 
will be reviewed here. In the process, it will also be attempted to clear some possible confusion. Also, quite a
few variations of local versions of fractional derivatives were introduced in the course of development.
It is planned to restrict to the results related to the original definition of LFD as given in~\cite{2KG1}
It is not the intent of this review to discuss and compare various versions of LFD though it can be an 
interesting topic in itself.

\section{Definitions}

The purpose of this section is to state the prerequisite definitions and also to introduce the LFD.
As we have stated earlier, we wish to modify the non-local fractional derivative to make it local.
We choose, for this purpose, the Riemann-Liouville derivative since, unlike Weyl derivative, it allows 
control over the lower limit and unlike Caputo derivative, it does not put extra smoothness conditions
on the function to be differentiated.

\begin{Def} The Riemann-Liouville fractional derivative of a function $f$ of order $q$ ($0<q<1$) is defined
as:
\begin{eqnarray}
D^q_xf(x') &=& \left\{\begin{array}{ll}
D^q_{x+}f(x'), & x'>x, \\
D^q_{x-}f(x'), & x' < x. 
 \end{array}\right. \nonumber \\
&=& \frac{1}{\Gamma(1-q)}
\left\{\begin{array}{ll}
\frac{d}{dx'}\int_x^{x'}f(t)(x'-t)^{-q} dt, & x'>x, \\
-\frac{d}{dx'}\int_{x'}^xf(t)(t-x')^{-q} dt, & x' < x. 
 \end{array}\right.
\end{eqnarray}

\end{Def}

Many a times it becomes necessary to study the local scaling behaviour of a function, especially in the case
of a fractal or a multifractal function. In such situations the non-local behaviour of the usual fractional derivatives
can be a hurdle in correctly characterizing the local properties of the functions.
Therefore, in~\cite{2KG1} we defined a \emph{local fractional derivative} (LFD) as:
\begin{Def} The local fractional derivative of order $q$ ($0<q<1$) of a function $f\in C^0: \mathbb{R} \rightarrow \mathbb{R}$
is defined as \[ {\cal{D}}^q f(x) = \lim_{x'\rightarrow x} D^q_x(f(x')-f(x)) \] if the limit exists in $\mathbb{R} \cup {\infty}$.

\end{Def}
Here, we have subtracted the value of the function $f$ at the point $x$ which is the point of interest. This washes
out the effect of the constant term making the definition independent of the origin or translationally invariant.
We have also introduced a limit which makes it explicitly local. Precursors to this definitions can be found in the
work of Hilfer~\cite{Hil1} and also in concurrent and independent work of Ben Adda~\cite{BA}.

Let us consider a simplest possible example: 

\begin{Ex} $f(x) = x^p$ where $0<p<1$ and $x\geq 0$. We would like to find out
the LFD of this function at $x=0$. Notice that $f(0)=0$. So we have
\begin{eqnarray} \nonumber
{\cal{D}}^qf(0) &=& {\lim_{x\rightarrow 0}{\frac{d^q(f(x)-f(0))}{d(x-0)^q}}} \\
&=& {\lim_{x\rightarrow 0}{\frac{d^q x^p }{dx^q}}} \nonumber \\
&=& {\lim_{x\rightarrow 0} \frac{\Gamma(p+1)}{\Gamma(p-q+1)} x^{p-q}} \nonumber \\
&=& {\left\{ \begin{array}{ll}
0 & q<p \; \mbox{or} \;\; q = p+n, \;\; n=1,2,3,...\\ \Gamma(p+1) & p=q \\ \infty & \mbox{otherwise} \\
 \end{array} \right. } \nonumber
\end{eqnarray}

At any other point $x = x_0 > 0$, $f(x)$ is differentiable with non-zero value of the derivative.
Hence it can be approximated locally as $f(x) = f(x_0) + f'(x_0) (x-x_0) + o(x-x_0)$. So the LFD
becomes
\begin{eqnarray} \nonumber
{\cal{D}}^qf(x_0) &=& {\lim_{x\rightarrow x_0}{\frac{d^q(f(x)-f(x_0))}{d(x-x_0)^q}}} \\
&=& f'(x_0) {\lim_{x\rightarrow x_0}{\frac{d^q (x-x_0) }{d(x-x_0)^q}}} \nonumber \\
&=&f'(x_0) {\lim_{x\rightarrow x_0} \frac{\Gamma(1+1)}{\Gamma(1-q+1)} (x-x_0)^{1-q}} \nonumber \\
&=& {\left\{ \begin{array}{ll}
0 & q<1 \;\mbox{or} \; q=2,3,4,...\\f'(x_0) \Gamma(2) & q=1 \\ \infty & \mbox{otherwise} \\
 \end{array} \right. } \nonumber
\end{eqnarray}
\label{ex:1}
\end{Ex}

One observes in this example that owing to the limiting procedure the LFD has a singular behaviour and 
the LFD is zero for orders smaller than certain \emph{critical order} and infinite for most of the orders above this order.
 This leads us to the
following definition:
\begin{Def}
The degree or the critical order of LFD of the continuous function $f$ at $x$ is defined as:
\[ q_c(x) = \sup\{q: {\cal{D}}^qf(x) \; \mbox{exists at}\; x \; \mbox{and is finite} \} \]
\end{Def}
In the example~\ref{ex:1}, the critical order is $p$ at $x=0$ and 1 at other values of $x$.

In order to generalize the definition~\cite{2KG2} to the orders beyond one we have to subtract the Taylor expansion around
the point of interest as follows:
\begin{Def}
The LFD of order $q$ ($N<q\leq N+1$) of a function $f\in C^0: \mathbb{R} \rightarrow \mathbb{R}$
is defined as \[ {\cal{D}}^q f(x) = \lim_{x'\rightarrow x} D^q_x\left(f(x')-\sum_{n=0}^N\frac{f^{(n)}(x)}{\Gamma(n+1)} (x'-x)^n\right) \] 
if the limit exists in $\mathbb{R} \cup {\infty}$.
\end{Def}
With this generalized definition, the critical order in the example~\ref{ex:1} gets modified.
Whereas, it has the same value $p$ at $x=0$, when $x>0$ it has value $\infty$ as the function
is analytic at all points $x>0$ and hence has a Taylor series expansion around any point $x>0$.

A generalization to a multivariable function has also been carried out~\cite{2KG3} in the following manner:
\begin{Def} 
We first define, for a function $f\in C^0: \mathbb{R}^n \rightarrow \mathbb{R}$,
\[ \Phi({\bf x},t) = f({\bf x}+{\bf v}t) - f({\bf x}) \;\;\; {\bf v} \in \mathbb{R}^n, \;\; t\in \mathbb{R}. \]
Then the directional-LFD of $f$ at ${\bf x}$ of order $q$ ($0<q<1$) in the direction ${\bf v}$
is defined as\[ {\cal{D}}^q_{\bf v} f(x) =  D^q_t \Phi({\bf y},t) |_{t=0} \]
if the limit exists in $\mathbb{R} \cup {\infty}$.

\end{Def}
It is interesting to notice that the limit we introduced to make the derivative local in the
previous definitions already exists in the definition of the ordinary directional derivative.
The generalization of this definition to higher order has been given in~\cite{BDG}.

In an interesting development, Chen et al.~\cite{CYZ} carried out careful analysis and, in particular,
proved the following lemma:
\begin{lem}
Let $f:(a,b)\rightarrow \mathbb{R}$ be continuous such that ${\cal{D}}_+^\alpha f(y)$ exists at some point
$y\in(0,1)$ then
\[
\lim_{h\rightarrow 0_+} \int_0^1 (1-t)^{-\alpha} \frac{f(ht+y)-f(y)}{h^\alpha} dt
\]
exists and
\begin{eqnarray}
{\cal{D}}_+^\alpha f(y) = lim_{h\rightarrow 0_+} \int_0^1 (1-t)^{-\alpha} \frac{f(ht+y)-f(y)}{h^\alpha} dt
\end{eqnarray}
\end{lem}

\section{Local fractional Taylor expansion}
A geometric interpretation is assigned to the LFD when we observe that it naturally appears
in a generalization of the Taylor expansion. If one follows the usual steps to derive the
Taylor expansion~\cite{2CJ} and replaces the ordinary derivative by the Riemann-Liouville
fractional derivative then what one ends up with is the local fractional Taylor expansion with
a remainder term.

Let
\begin{eqnarray}
F(x,x'-x;q) = {D^q_x(f(x')-f(x))}.
\end{eqnarray}
It is clear that
\begin{eqnarray}
{\cal{D}}^qf(x)=F(x,0;q)
\end{eqnarray}
Now, for $0<q\leq 1$,
\begin{eqnarray}
f(x')-f(x)&=& {D^{-q}_x}{D^{q}_x}(f(x')-f(x)) \nonumber\\
& =& {1\over\Gamma(q)} \int_0^{x'-x} {F(x,t;q)\over{(x'-x-t)^{-q+1}}}dt\\
&=& {1\over\Gamma(q)}[F(x,t;q) \int (x'-x-t)^{q-1} dt]_0^{x'-x} \nonumber\\
&&\;\;\;\;\;\;\;\;+ {1\over\Gamma(q)}\int_0^{x'-x} {dF(x,t;q)\over{dt}}{(x'-x-t)^q\over{q}}dt
\end{eqnarray}
provided the last term exists. Thus
\begin{eqnarray}
f(x')-f(x)&=& {{\cal{D}}^qf(x)\over \Gamma(q+1)} (x'-x)^q \nonumber\\
&&\;\;\;\;\;\;\;\;+ {1\over\Gamma(q+1)}\int_0^{x'-x} {dF(x,t;q)\over{dt}}{(x'-x-t)^q}dt\label{taylor}
\end{eqnarray}
i.e.
\begin{eqnarray}
f(x') = f(x) + {{\cal{D}}^qf(x)\over \Gamma(q+1)} (x'-x)^q + R_q(x',x) \label{taylor2}
\end{eqnarray}
where $R_q(x',x)$ is a remainder given by
\begin{eqnarray}
R_q(x',x) = {1\over\Gamma(q+1)}\int_0^{x'-x} {dF(x,t;q)\over{dt}}{(x'-x-t)^q}dt
\end{eqnarray}
Equation (\ref{taylor2}) is a fractional Taylor expansion of $f(x)$ involving
only the lowest and the second leading terms. Using the general definition
of LFD and following similar steps one arrives at the fractional Taylor
expansion for $N<q\leq N+1$ (provided ${\cal{D}}^q$ exists and is finite), given by,
\begin{eqnarray}
f(x') = \sum_{n=0}^{N}{f^{(n)}(x)\over{\Gamma(n+1)}}(x'-x)^n
 + {{\cal{D}}^qf(x)\over \Gamma(q+1)} (x'-x)^q + R_q(x',x) \label{taylorg}
\end{eqnarray}
where 
\begin{eqnarray}
R_q(x',x) = {1\over\Gamma(q+1)}\int_0^{x'-x} {dF(x,t;q,N)\over{dt}}{(x'-x-t)^q}dt
\end{eqnarray}
We note that the local fractional derivative (not just fractional derivative)
as defined above
 provides
the coefficient $A$ in the approximation
of $f(x')$ by the function $f(x) + A(x'-x)^q/\Gamma(q+1)$, for $0<q<1$, 
in the vicinity of $x$. This 
generalizes the geometric interpretation of derivatives in terms of `tangents'.
It can be shown~\cite{AC} that if the LFD in equation~(\ref{taylorg}) exists
then the remainder term $R_q(x',x)$ goes to zero as $x\rightarrow x'$.
\begin{Ex}\label{ex:2}
Let us consider a function $f(x) = ax^\alpha + b x^\beta$ where $x\geq 0$ and $0<\alpha < \beta < 1$
and study its local fractional Taylor expansion at $x=0$. If $q< \alpha$, then ${\cal{D}}^qf(x)=0$
and the remainder term evaluates to $ax^\alpha + b x^\beta$. But if $q=\alpha$, the critical order of $f$, then the
second term in the equation~(\ref{taylorg}) is finite and equal to $ax^{\alpha}$ and the remainder
term yields $bx^{\beta}$.

\end{Ex}

The example illustrates the importance of working at the critical order of the function.
Osler, in ref.~\cite{2Osl}, has constructed fractional Taylor
series using usual (not local in the sense above) fractional derivatives. 
His results are, however, applicable to analytic functions and cannot be 
used for non-differentiable scaling functions directly. Furthermore, Osler's
formulation involves terms with negative $q$ also and hence is not suitable
for approximating schemes. Finally, the local fractional Taylor series has also been
extended~\cite{BDG} to multivariable functions.

\section{Local fractional differentiability}
Fractional differentiability of functions can be studied using any definition
of the fractional derivative which would signify the existence of that derivative
for a given function. Considering \emph{local} fractional differentiability has
an added advantage that it preserves the local property of differentiability
in the ordinary calculus. Thus it becomes unnecessary to worry about the definition
of the function away from the point of interest. We refer the reader to reference~\cite{LZ}
for a recent review of different notions of differentiability and their interrelation.

In~\cite{2KG1}, local fractional differentiability of Weierstrass' everywhere continuous
but nowhere differentiable function was studied. A form of this function is given by
\begin{eqnarray}
W_{\lambda}(t) = \sum_{k=1}^{\infty} {\lambda}^{(s-2)k} 
\sin{\lambda}^kt,\;\;\;\;
\lambda>1\;\;\;1<s<2, \label{eq:Wsp}
\end{eqnarray}
where $\lambda > 1$ and $1<s<2$. Note that $W_{\lambda}(0)=0$. The graph of this function
is fractal with dimension $s$ and the H\"older exponent at every point of this function
is $2-s$. It was shown in~\cite{2KG1} that the Weierstrass' function is locally fractionally
differentiable up to order $2-s$ at every point and the LFD does not exist anywhere for orders
greater that $2-s$.

In fact, we proved a general result~\cite{2KG1} showing the connection between the degree of LFD
(critical order) and the H\"older exponent. In~\cite{2KG1} we used the  definition of the \emph{pointwise}
H\"older exponent but it turns out that the right way is to use the \emph{local} H\"older exponent. This
was pointed out in~\cite{KLV}. The following theorem was proved there:
\begin{The}
Let $f$ be a continuous function in $L^2$. Then $q_c(f,x_0) = \alpha_l(f,x_0)$ where
$\alpha_l(f,x_0)$ is the local H\"older exponent of $f$ at $x_0$.

\end{The} 

The local fractional differentiability of a multifractal function, in which the H\"older
exponent varies from point to point, was also studied~\cite{2KG1}.

As of now, we do not know
any example of a continuous everywhere but 
nowhere differentiable function for which the LFD 
exist at the critical
order. In fact, it was suspected by the author during his thesis that no such function might
exist. This is because, a close inspection of such functions shows oscillations of the
log-periodic type at every point of these functions which possibly leads to the nonexistence
of LFD at the critical point. We have demonstrated such oscillations in the case of devil's
staircase~\cite{KK}. But this should not be a serious hurdle as the log-periodic oscillations
only reflect the lacunarity in the fractal functions. The lacunarity has been discussed
at length by Mandelbrot in~\cite{1Man}. But the critical order of the LFD correctly characterizes
the H\"older exponent and hence the dimension of the fractal functions. This means,
there are two possible future directions of exploration: one is to assume that the LFD
exists at the critical order and develop its applications and the other is to try to modify the definition of
the LFD. In the following sections we take the first approach.

\section{Properties of LFD}
In this section, we review some of the properties of LFD as studied by different authors.
Papers in questions are mainly~\cite{AC},~\cite{BDG} and~\cite{CCC}. In~\cite{AC} Ben Adda and Cresson 
and in~\cite{BDG}, Babakhani and Daftardar-Gejji
proved quite a few properties of LFD rigorously. The first one is a Leibnitz rule for a 
product of two functions one of which is smooth and other could be non-differentiable.
The following theorem was proved in~\cite{BDG}:
\begin{The}
Let $f(x)$ be continuous on $[a,b]$ and ${\cal{D}}_+^\alpha f(a)$, ${\cal{D}}_-^\alpha f(b)$ and ${\cal{D}}_\pm^\alpha f(x)$
exists for every $x\in(a,b)$. If further $\phi(x)\in C^3[a,b]$, then for $0<\alpha<1$
\[ {\cal{D}}_+^\alpha\left( (\phi f)(a)\right) = \phi(a){\cal{D}}_+^\alpha f(a), \]
\[ {\cal{D}}_-^\alpha\left( (\phi f)(b)\right) = \phi(b){\cal{D}}_-^\alpha f(b), \]
\[ {\cal{D}}_\pm^\alpha\left( (\phi f)(x)\right) = \phi(x){\cal{D}}_\pm^\alpha f(x). \]
\end{The}
They also extended the result to the case $n < \alpha < n+1$.
In~\cite{CCC}, Carpinteri et al. used a more general rule for the LFD of a product of two functions.
They considered the case when both functions were non-differentiable. If $f$ and $g$ are two functions
having the same H\"older exponent, say $\alpha$, then
\[ {\cal{D}}^\alpha \left(f(x)g(x)\right)= f(x){\cal{D}}^\alpha g(x) + g(x){\cal{D}}^\alpha f(x). \]
This was already proved by Ben Adda and Cresson in~\cite{AC}. They also proved the following formula for
the LFD of division of two functions when the LFD of individual functions exist:
\[ {\cal{D}}^\alpha \left(\frac{f(x)}{g(x)}\right)= \frac{g(x){\cal{D}}^\alpha f(x) + f(x){\cal{D}}^\alpha g(x)}{g^2(x)}, \]
where $g(x)\neq 0$.

The next property studied in~\cite{BDG} is the chain rule, that is, the LFD of composite function.
The following theorem was proved there.
\begin{The}
Let $h:[a,b]\rightarrow \mathbb{R}$ be a function of class $C^{n+3}$, $f$ be a function of class $C^n$ on
$h[a,b]$, and ${\cal{D}}_\pm^{\alpha - n}\left[ f^{(n)}\left(h(x)\right)\right]$ exists. Then
\[ {\cal{D}}_\pm^{\alpha}\left[ f\left(h(x)\right)\right] = \left( \frac{dh}{dx} \right)^n
{\cal{D}}_\pm^{\alpha - n}\left[ f^{(n)}\left(h(x)\right)\right] \]
where $n < \alpha < n+1$, $n\in \mathbb{N} \cup \{0\}$.
\end{The}
The reader is referred to~\cite{AC} for an interesting generalization of the chain rule.

\section{Local fractional differential equations}
Once we have the LFD it is natural to ask if we can write and solve equations in terms of this
operator. The simplest such equation will have the form
\begin{eqnarray}\label{eq:lfde}
{\cal{D}}^\alpha f(x) = g(x)
\end{eqnarray}
where $g(x)$ is some known function and $f(x)$ is an unknown function to be found out
and $0<\alpha<1$. The immediate
question that arises is what is the class of function $g(x)$ for which the solution exist. This question
will be considered in this section. A formal solution to the above equation can be written down
by generalizing the Riemann sum as follows:
\begin{eqnarray}\label{eq:solfde}
f(x)
=\lim_{N\rightarrow \infty} \sum_{i=0}^{N-1} {(x_{i+1}-x_i)^\alpha
\over\Gamma(\alpha+1)} g(x^*)
\end{eqnarray}
where $x_i \leq x^* \leq x_{i+1}$. This question was first asked in~\cite{2KG5} where the formal solution
in the form above was also discussed. The factor $(x_{i+1}-x_i)^\alpha / \Gamma(\alpha+1) $ was motivated by 
the local Taylor expansion. Later, in~\cite{2KG4,2KG6}, the same factor was written alternatively as an RL integral of constant function 1
of order $\alpha$, that is, $_{x_i}I^{\alpha}_{x_{i+1}} 1 $. This allowed us to define a differential of fractional order and
gave another motivation for the factor $\Gamma(\alpha+1)$ in the denominator of equation~(\ref{eq:solfde}).
We denoted the differential of fractional order as $d^{\alpha}x$, in line with the notation $d^3x$ for the volume element,
and introduced the notation
\begin{eqnarray}\label{eq:symbol}
\int g(x) d^{\alpha}x
\end{eqnarray}
for the RHS of equation~(\ref{eq:solfde}) calling it a "fractal integral" or "local fractional integral". 
So, ~(\ref{eq:symbol}) is only a symbol for the RHS of equation~(\ref{eq:solfde}).

It is easy to see that the solution~(\ref{eq:solfde}) does not exist if
$g(x)$ is a continuous function.
\begin{Prop}
The solution (\ref{eq:solfde}) of the local fractional differential equation (LFDE) (\ref{eq:lfde})
diverges if $g(x)$ is a continuous function.
\end{Prop}
{\bf \it Proof:} If $g(x)$ is a continuous function then there exists an interval, say $[\delta,\gamma]$,
such that $g(x)$ is non-zero positive on this interval. That is, there exists $\epsilon > 0$ such that
$g(x) > \epsilon$ on this interval. In this interval, we have
\begin{eqnarray}\label{eq:solfdep}
f(x)
&=&\lim_{N\rightarrow \infty} \sum_{i=0}^{N-1} {(x_{i+1}-x_i)^\alpha
\over\Gamma(\alpha+1)} g(x^*) \nonumber \\
&\geq & \lim_{N\rightarrow \infty} \sum_{i=0}^{N-1} {(x_{i+1}-x_i)^\alpha
\over\Gamma(\alpha+1)} \epsilon \nonumber \\
&=& \lim_{N\rightarrow \infty} N \times \left( \frac{\gamma - \delta}{N}\right)^\alpha \epsilon \nonumber \\
&=& \lim_{N\rightarrow \infty} N^{1-\alpha} (\gamma - \delta)^\alpha \epsilon \nonumber \\
&=& \infty. \nonumber
\end{eqnarray}
Hence the result.

This implies that the LFDE (\ref{eq:lfde}) will have a solution only if $g(x)$ is a discontinuous function.
This also means that any function can not have continuous LFD in any interval as argued in~\cite{LDE}.
That is $g(x)$ can be of two types or their combination: (i) the function $g(x)$ is an indicator function of some sparse set (like
the Cantor set) such that
the number of terms in the sum of (\ref{eq:solfde}) grow sub-linearly and hence the limit is finite,
and (ii) the function $g(x)$ alternates between positive and negative values in any small interval
giving rise to cancellations among the terms of the sum in~(\ref{eq:solfde}) leading to convergent
limit. There is a classic Conway base 13 function which is an example of a function which is discontinuous
in any given interval. Another possible example, perhaps of practical importance, would be white noise. 
In the following two subsections
we discuss each case in more detail.

It is 
interesting to note that though such integrals arise naturally
in our formalism as inverse of LFD, Mandelbrot already
in~\cite{1Man} has suggested studying such integrals using nonstandard
analysis which extends the real number system to include infinite
and infinitesimally small number. However, as is made clear in
the following, since we restrict $g(x)$ to two classes of 
physically meaningful functions for which this integral is finite
it obviates the need to use the nonstandard analysis

\subsection{$g(x)$ as indicator function of a Cantor set}
As discussed above, the solution to the LFDE~(\ref{eq:lfde}) is seen to exist
if $g(x)$ is an indicator function of a fractal set. That is, if $C$ is a Cantor set
then $g(x)=1$ if $x\in C$ and $g(x)=0$ otherwise. We denote this by $g(x)=1_C(x)$.
We now proceed to see that the solution with
initial condition $f(0)=0$ exists if $\alpha = \mbox{dim}_HC$.
In this case the above Riemann sum~(\ref{eq:solfde}) takes the form 
\begin{eqnarray}
f(x)
=\lim_{N\rightarrow \infty} \sum_{i=0}^{N-1} {(x_{i+1}-x_i)^\alpha
\over\Gamma(\alpha+1)} F_C^i
\equiv { P_C(x)\over\Gamma(\alpha+1)}, \label{eq:PCt}
\end{eqnarray}
where $x_i$ are  subdivision points of the interval $[x_0=0, x_N=x]$ and
$F_C^i$ is a flag function which takes value 1 if the interval $[x_i,x_{i+1}]$
contains a point of the set $C$ and 0 otherwise. 
Now if we divide the interval $[0,x]$ into equal sub-intervals and denote
$\Delta=\Delta_i = x_{i+1}-x_i$ then we have $P_C(x)= \Delta^\alpha \sum F_C^i$.
But $\sum F_C^i$ is of the order of  $N^{-\alpha}$. Therefore
$P_C(x)$ satisfies the bounds 
$ax^\alpha \leq P_C(x) \leq bx^\alpha$ where $a$ and  $b$
are suitable positive constants. 
Note that $P_C(x)$ is a Lebesgue-Cantor 
(staircase) function. Since the function $1_C(x)$
is zero almost everywhere the function $P_C(x)$ is constant almost everywhere.
As is clear from the equation~(\ref{eq:PCt}), it rises only at points where $1_C(x)$ is non-zero.

Of course, this is the simplest example considered to elucidate the basic principle.
One can consider different forms for $g(x)$, such as any function $h(x)$ multiplied by
the indicator function of the Cantor set $1_C(x)$. Such an equation also will have
solutions. Complete theory of these equations is still to be worked out. Some more 
examples of this kind will be considered in the next section discussing applications.

\subsection{$g(x)$ as white noise}
The second class, which we are going to consider in this subsection, 
consists of rapidly oscillating
functions which oscillate around zero in any small interval. 
This type of equation was introduced in~\cite{KK1} though its 
mathematical details are still to be worked out.
These oscillations then would result in cancellations again giving
rise to a finite solution.
A realization of the white noise
is one example in this class of functions. This immediately prompts us to
consider a generalization of the Langevin equation which involves LFD
and $g(x)$ is chosen as white noise.

So we consider a generalization of the 
Langevin equation~\cite{Risken1989} in high friction
limit where one neglects the acceleration term
and replaces the first derivative term by the LFD to arrive at
\begin{eqnarray}\label{eq:lengevin}
{\cal{D}}^{\alpha}x(t) = \zeta(t), 
\end{eqnarray}
where $<\zeta(t)>=0$ and $<\zeta(t)\zeta(t')>=\delta(t-t')$
the Dirac delta function.
The solution of the above equation follows from Eq.~(\ref{eq:solfde})
\begin{eqnarray}
x(t) = \lim_{N\rightarrow\infty} \sum_{i=0}^{N-1} \zeta_i
{(t_{i+1}-t_i)^\alpha\over\Gamma(\alpha+1)}.
\end{eqnarray}
Heuristically it can be seen that
\begin{eqnarray}
<x(t)> &=& \lim_{N\rightarrow\infty} 
{N^{-\alpha}\over\Gamma(\alpha+1)} N^{1/2} 
\end{eqnarray}
and therefore the average is zero if $\alpha > 1/2$ and it does not
exist if $\alpha < 1/2$. This indicates that the above process is
of  L\'evy type with index $2\alpha$ for $\alpha < 1$.

\section{Applications of LFDE}
One would immediately wonder about the applications of such equations involving
LFDs. We consider two of them over here. The first one was considered in~\cite{2KG5}
which considers a local fractional diffusion equation and the second one was
introduced by Carpinteri and Cornetti~\cite{CC} in which they generalized the relation
between the stress and strain for fractal media using the LFDE.

\subsection{Local fractional diffusion equation}
In this subsection we consider the local fractional diffusion equation.
The equation can be derived systematically starting from the Chapman-Kolmogorov
condition and making use of the local fractional Taylor expansion. The reader
is referred to~\cite{2KG5} for more details. Here we consider the equation and
its solution.
\begin{eqnarray}
{\cal{D}}_t^{\alpha}W(x,t) 
&=& {\Gamma(\alpha + 1)\over 4} \chi_C(t) {\partial^2\over{\partial x^2}}W(x,t)
\label{eq:ex} 
\end{eqnarray}
We note that even though the variable $t$ is taking all real positive values
the actual evolution takes place only for values of $t$ in the fractal set $C$.
The solution of equation~(\ref{eq:ex}) can easily be obtained as
\begin{eqnarray}
W(x,t) 
&=& P_{t-t_0}W(x,t_0)
\end{eqnarray}
where
\begin{eqnarray}
P_{t-t_0}
&=&  \lim_{N\rightarrow \infty} \prod_{i=0}^{N-1} \big[ 1 + {1\over 4} (t_{i+1}-t_{i})^\alpha F_C^i {\partial^2\over{\partial
x^2}}\big].
\end{eqnarray}
The above product converges because except for terms for which $F^i_C=1$  
(which are of order $N^{\alpha}$) all others take value 1.
It is clear that for $t_0 < t' < t$
\begin{eqnarray}
W(x,t) = P_{t-t'}P_{t'-t_0}W(x,t_0)
\end{eqnarray}
and $P_t$ gives rise to a semigroup evolution.
Using equation~(\ref{eq:PCt}) it can be 
easily seen that
\begin{eqnarray}
W(x,t) = e^{{P_C(t)\over 4}{\partial^2\over{\partial
x^2}}}W(x,t_0=0).
\end{eqnarray}
Now  choosing the initial distribution $W(x,0) = \delta(x)$ and using the
 Fourier representation
of delta function, we get the solution
\begin{eqnarray}
W(x,t) &=& {1\over\sqrt{\pi P_C(t) }}e^{ - x^2\over{P_C(t)}} \label{eq:sol}
\end{eqnarray}
Consistency of the equation~(\ref{eq:sol}) can easily be checked by 
directly substituting this in Chapman-Kolmogorov
equation.
We note that this solution satisfies the bounds
\begin{eqnarray}
{1\over\sqrt{\pi bt^\alpha }}e^{ - x^2\over{bt^\alpha}} \leq 
W(x,t)  \leq
{1\over\sqrt{\pi at^\alpha }}e^{ - x^2\over{at^\alpha}}
\end{eqnarray}
for some $0<a<b<\infty$.
This is a model solution of a subdiffusive behaviour. It is clear that
when $\alpha=1$ we get back the typical solution of the ordinary
diffusion equation, which is $(\pi t)^{-1/2}\exp(-x^2/t)$.

\subsection{Relation between stress and strain for fractal media}
Carpinteri and Cornetti~\cite{CC} used the concept of LFD to write 
down a relation between the strain and the displacement when the 
strain is localized on the fractal set. They proposed
\begin{eqnarray}\nonumber
\epsilon^*(x) = {\cal{D}}^\alpha u(x)
\end{eqnarray}
where $\epsilon^*(x)$ is a renormalised strain and $u(x)$ is the displacement.
This also offers a physical interpretation to LFD. This formalism is useful
in studying the structural properties of concrete-like or disordered materials.
In~\cite{CCK}, we calculated resultant of a stress distribution and its moment
when the stress is distributed over a fractal set. Further use of LFD and the fractal
integral for the principle of virtual work and to generalize the constitutive 
equations of elasticity has been made in~\cite{CCC, CCC1}.

\section{Conclusions}
We have reviewed developments emanating from and related to the notion of
local fractional derivative introduced in~\cite{2KG1}. The concept has received
wide attention and led to rich variety of works. It was not possible to review
all of them here. This article was restricted to the developments directly
related to the definition of LFD introduced originally in~\cite{2KG1}. Also,
there seem to be some misconceptions leading to some erroneous results.
It is hoped that this review will help to dispel some doubts.

The central idea was to appropriately modify the non-local fractional derivatives
so as to be able to characterize local scaling behaviour which gave rise to LFD.
The LFD so defined happened to arise naturally in the local Taylor expansion
which meant that there was something more to the construction of the LFD though
it involved seemingly artificial steps of subtracting the value of the function
at the point and an additional limiting procedure. It was important to realize
the LFD as a standalone operator even if its action on the function is "singular"
in the sense that the result is zero for most of the orders and hence pursue its
properties and applications. Several authors have carried out substantial work
in this direction. The LFD gave rise to the notion of local fractional differentiability
and the relation between the order of differentiability and the local H\"older exponent.
Considering LFD as an operator in its own right leads to questions of its inverse
and also equations involving LFD leading to local fractional differential equations
and local fractional integrals or fractal integrals which is supposed to invert
the LFDs. As a result, local fractional diffusion equation, an equation in terms of
LFD to describe the relation between stress and strain in fractal media, local
fractional stochastic differential equation are some of the outcomes these series
of explorations.

Clearly, lot needs to be done. On the one had, putting the formalism on rigorous
mathematical footing is something that needs to be pursued vigorously. Also,
developing more applications in diverse fields will give impetus to the efforts
put in the development of the theory.

\textbf{Acknowledgements}\\
I would like to thank Vishwesh Vyawahare and Mukesh Patil for inviting me to the workshop on Applications of
Fractional Calculus in Engineering which they organized in Navi Mumbai, India in March 2012, Guo-Cheng Wu and Wen
Chen for inviting to be part of FDA12 in Nanjing, China in May 2012 and Varsha Daftardar-Gejji for involving
me in the Workshop on Fractional Calculus: Theory and Applications held in Pune, India in November 2012. These
three events in quick succession has brought me back to the field and made me realize that there is a lot to
catch up with!

\end{document}